\newcounter{myctr}
\def\myitem{\refstepcounter{myctr}\bibfont\noindent\ifnum\themyctr>9\else\phantom{0}\fi\hangindent17pt\themyctr.\enskip}

\documentclass{suddenchangeofIP}
\usepackage{hyperref}
\usepackage[super,sort,compress]{cite}
\usepackage{mathrsfs}
\usepackage{amsfonts}
\usepackage{amssymb}
\usepackage{amsmath}
\usepackage{epsfig}
\usepackage{makeidx}
\usepackage{setspace}
\usepackage{rotating}
\usepackage{dcolumn}
\usepackage{bm}
\usepackage{bbold}
\usepackage{graphicx}
\usepackage{subfigure}
\begin{document}

\catchline{}{}{}{}{}

\title{SUDDEN CHANGE OF INTERFEROMETRIC POWER FOR X SHAPE STATES}

\author{Dian Zhu}

\address{Theoretical Physics Division, Chern Institute of Mathematics, Nankai University, Tianjin 300071, China}

\author{Fu-Lin Zhang}

\address{Department of Physics, School of Science, Tianjin University, Tianjin 300072, China \\
flzhang@tju.edu.cn}

\author{Jing-Ling Chen}

\address{Theoretical Physics Division, Chern Institute of Mathematics, Nankai University, Tianjin 300071, China \\
chenjl@nankai.edu.cn}

\maketitle


\begin{abstract}
Quantum interferometric power (IP) is a discordlike measure.
We study the dynamics of IP for two-qubit X shape states under different noisy environments.
Our study shows that IP exhibits sudden change, and one side quantum channel is enough for the occurrence of a sudden change of IP.
In particular, we show that the initial state having no sudden change of quantum discord exhibits a sudden change of IP under the dynamics of amplitude noise, but the converse is not true.
Besides, we also investigate the dynamics of IP under two different kinds of composite noises.
Our results also confirm that sudden change of IP occurs under such composite noises.
\end{abstract}



\markboth{Authors' Names}
{Instructions for Typing Manuscripts (Paper's Title)}

\section{Introduction}	

Quantum correlations characterize the quantum feature of bipartite or multipartite system.\cite{Nielsen2000}
For a long time, entanglement was considered to be the only quantum correlation that was useful for quantum information processing.
However, it is realized that there exists another quantum correlations weaker than entanglement called quantum discord,\cite{PRL88017907,JPA346899}
which can be found in separable mixed states and may play an important role in quantum information processing.
For instance, quantum discord can be found to be present in the deterministic quantum computation with one qubit (DQC1) while there is no entanglement.\cite{PRL100090502,PRL101200501}
In the past two decades, a great deal of attentions have been received for quantum discord.\cite{PRL100090502,PRA67012320,PRA71062307,PRA76032327,PRA77042303,PRA80024103,PRA80022108,PRA80044102,
PRA80052304,PRA81042105,PRA82069902,PRA84042313,PRA83022321,PRA86032110,PRL100050502,PRL102250503,PRL102100402,PRL105190502,PRL104200401,PRL107140403,PRA85032318,PRA86012312,PST153014036}

One interesting and challenging topic in quantum information theory is characterizing and quantifying such nonclassical correlation.
Recently, several discordlike quantum correlation measures were proposed and studied from different perspectives, one of which was interferometric power (IP).\cite{PRL110240402,PRL112210401}
Based on quantum fisher information, the definition of IP naturally quantified the guaranteed sensitively of the probe states in some interferometric devices.\cite{PRL112210401}
As one of the discordlike measures, IP has some basic properties, such as invariance under local unitary operations, and nonnegativity for states of classical correlation.\cite{PRA97032326}

We already know that {\it sudden change}\cite{PRA80044102,PRL104200401} is a peculiar dynamical behavior of the dynamics of quantum discord for two-qubit systems passing through quantum noise channels.
But there are no investigations to show that the sudden change of IP exists for two-qubit systems under the same cases.
In this paper, we investigate the dynamics of IP for two-qubit X shape states under several different kinds of quantum noises.
Our results show that compared to depolarizing noise, IP exhibits sudden change under amplitude noise and phase noise acting on the first qubit of the two-qubit system.
In particular, our findings show that the condition of the sudden change of IP differs from quantum discord under amplitude noise acting on the first qubit of the two-qubit system.
Our studies also show that IP exhibits sudden change under the two different kinds of dephasing noise acting on both two-qubit systems.

\section{Quantum IP as the discordlike measure of X shape states}

The concept of IP of quantum states has been introduced explicitly in Ref.~\citen{PRL112210401}. 
As one of the measure of discordlike quantum correlations, IP provides a computable way of quantum discord quantification.
For a bipartite system $\rho_{AB}$, the IP is defined as the minimum of the Quantum Fisher Information (QFI)\cite{PRL723439}
by taking over all the local Hamiltonians $H_A$ acted only on the subsystem A,
\begin{equation} \label{IPrhoAB}
\mathcal{IP}(\rho_{AB}) = \dfrac{1}{4} \min_{H_A} \mathcal{F}(\rho_{AB},H_A),
\end{equation}
where
\begin{equation}
\mathcal{F}(\rho_{AB},H_A) = 4 \sum_{i\ne j;q_i +q_j \ne 0} \dfrac{(q_i -q_j)^2}{q_i +q_j} 
\left| \langle\psi_1|H_A \otimes\mathbb{1}_B |\psi_j\rangle \right|^2
\end{equation}
with $\{ q_i, |\psi_i\rangle \}$ being the eigenvalues and eigenvectors of $\rho_{AB}$, respectively.
For the case of subsystem A being a qubit, one can take the set of Hamiltonians $H_A = \vec{n}\cdot\vec{\sigma}$ with $\vec{n}=1$ and $\vec{\sigma} = (\sigma_x,\sigma_y,\sigma_z)$.
Then Eq.~(\ref{IPrhoAB}) can be reduced to a closed formula
\begin{equation}
\mathcal{IP}(\rho_{AB}) = \zeta_{min}[M],
\end{equation}
where $\zeta_{min}[M]$ is the smallest eigenvalues of the matrix $M$ with elements\cite{PRL112210401}
\begin{equation} \label{Melement}
M_{m,n} = \dfrac{1}{2} \sum_{i,l;q_i+q_l \ne 0}\dfrac{(q_i-q_l)^2}{q_i+q_l}
\langle\psi_i|\sigma_m \otimes \mathbb{1} |\psi_l\rangle \langle\psi_l| \sigma_n \otimes \mathbb{1} |\psi_i\rangle.
\end{equation}
One can note that IP has the following properties:
(i) $\mathcal{IP}(\rho_{AB})$ equals to zero if $\rho_{AB}$ is a classical state (with respect to $A$);
(ii) $\mathcal{IP}(\rho_{AB})$ is invariant under local unitary operation;
(iii) $\mathcal{IP}(\rho_{AB})$ is monotonically decreasing under local completely positive and trace preserving maps on subsystem $B$;
(iv) $\mathcal{IP}(\rho_{AB})$ reduces to a measure of entanglement if $\rho_{AB}$ is a pure state.\cite{PRL112210401}
These properties imply that $\mathcal{IP}(\rho_{AB})$ is a proper measure of discord-type correlation.

Let us consider the following X shape states of two-qubit system,\cite{PRA83022321}
\begin{equation}  \label{Xstate}
\rho_{AB} = \frac{1}{4} (\mathbb{1}\otimes\mathbb{1} + \vec{r}\cdot\vec{\sigma}\otimes\mathbb{1} + \mathbb{1}\otimes\vec{s}\cdot\vec{\sigma} +\sum_{j=1}^{3} c_j \sigma_j \otimes \sigma_j),
\end{equation}
where $\mathbb{1}$ is the $2 \times 2$ identity matrix, $\vec{r} = (0,0,r)$, $\vec{s} = (0,0,s)$, $c_j \in \mathbb{R}$, and $\sigma_j (j=1,2,3)$ are the Pauli matrices.
It is easy to see that Eq.~(\ref{Xstate}) reduces to Bell-diagonal states when $r=s=0$.

The form of IP of $\rho_{AB}$ can be given by Eq.~(\ref{Melement}), that is
\begin{equation} \label{IP2}
\mathcal{IP}(\rho_{AB}) = \min\{ M_{11},M_{22},M_{33} \},
\end{equation}
where
\begin{equation} \label{M123}
  \begin{split}
     M_{11} & = \frac{(\lambda_1-\lambda_3)^2}{\lambda_1+\lambda_3} \cdot \frac{(x_1+y_1)^2}{(1+x_1^2)(1+y_1^2)} +
                \frac{(\lambda_1-\lambda_4)^2}{\lambda_1+\lambda_4} \cdot \frac{(x_1+y_2)^2}{(1+x_1^2)(1+y_2^2)} \\
            & + \frac{(\lambda_2-\lambda_3)^2}{\lambda_2+\lambda_3} \cdot \frac{(x_2+y_1)^2}{(1+x_2^2)(1+y_1^2)} +
                \frac{(\lambda_2-\lambda_4)^2}{\lambda_2+\lambda_4} \cdot \frac{(x_2+y_2)^2}{(1+x_2^2)(1+y_2^2)}, \\
     M_{22} & = \frac{(\lambda_1-\lambda_3)^2}{\lambda_1+\lambda_3} \cdot \frac{(x_1-y_1)^2}{(1+x_1^2)(1+y_1^2)} +
                \frac{(\lambda_1-\lambda_4)^2}{\lambda_1+\lambda_4} \cdot \frac{(x_1-y_2)^2}{(1+x_1^2)(1+y_2^2)} \\
            & + \frac{(\lambda_2-\lambda_3)^2}{\lambda_2+\lambda_3} \cdot \frac{(x_2-y_1)^2}{(1+x_2^2)(1+y_1^2)} +
                \frac{(\lambda_2-\lambda_4)^2}{\lambda_2+\lambda_4} \cdot \frac{(x_2-y_2)^2}{(1+x_2^2)(1+y_2^2)}, \\
     M_{33} & = \frac{(\lambda_1-\lambda_2)^2}{\lambda_1+\lambda_2} \cdot \frac{(x_1 x_2-1)^2}{(1+x_1^2)(1+x_2^2)} +
                \frac{(\lambda_3-\lambda_4)^2}{\lambda_3+\lambda_4} \cdot \frac{(y_1 y_2-1)^2}{(1+y_1^2)(1+y_2^2)}.
  \end{split}
\end{equation}
Here $\lambda_i \geq 0$ are the eigenvalues of $\rho_{AB}$,
$x_1 = \frac{r-s-2(\lambda_2-\lambda_1)}{c_1+c_2}$,
$x_2 = \frac{r-s+2(\lambda_2-\lambda_1)}{c_1+c_2}$,
$y_1 = \frac{r+s-2(\lambda_4-\lambda_3)}{c_1-c_2}$ and
$y_2 = \frac{r+s+2(\lambda_4-\lambda_3)}{c_1-c_2}$.
From the above expressions, one can find that $M_{11} \geq M_{22}$ (or $M_{11} < M_{22}$) if $|c_1| \leq |c_2|$ (or $|c_1| > |c_2|$).
Hence, the IP of $\rho_{AB}$ can be rewritten as
\begin{equation}\label{IP3}
  \mathcal{IP}(\rho_{AB}) = \left\{ 
    \begin{array}{cc}
      \min\{ M_{22},M_{33} \}, & \text{if} |c_1| < |c_2|, \\
      \min\{ M_{11},M_{33} \}, & \text{if} |c_1| > |c_2|.
    \end{array}
\right.
\end{equation}

\section{Sudden change of IP under one side quantum channel}

In this section, based on the analytical formula of IP which has been given above, we can study the dynamics of IP for two-qubit system over three kinds of quantum noises acted on the first qubit: amplitude noise, phase noise and depolarizing noise.
For simplicity, we consider the Bell-diagonal states as the initial states of two-qubit system,\cite{PRA77042303}
\begin{equation}\label{BState}
  \rho = \frac{1}{4} (\mathbb{1}\otimes\mathbb{1} + \sum_{i=1}^{3} c_i \sigma_i\otimes\sigma_i).
\end{equation}

\subsection{Amplitude noise}

Amplitude damping, or amplitude noise, which is used to characterize spontaneous emission, describes the energy dissipation from a quantum system.
One can consider a two-qubit system where the first qubit is through this quantum channel.
Then the Kraus operators for the whole system are given by\cite{PRL93140404}
\begin{equation}
  \begin{split}
     K_{1a} &= \left(
     \begin{array}{cc}
       \eta & 0 \\
         0  & 1 
     \end{array}
      \right) \otimes
      \left(
     \begin{array}{cc}
         1 & 0 \\
         0  & 1 
     \end{array}
      \right), \\
     K_{2a} &= \left(
     \begin{array}{cc}
              0        & 0 \\
      \sqrt{1-\eta^2}  & 1 
     \end{array}
      \right) \otimes
      \left(
     \begin{array}{cc}
         1 & 0 \\
         0  & 1 
     \end{array}
      \right), \\
  \end{split}
\end{equation}
where $\eta = e^{-\frac{\tau t}{2}}$, $\tau$ is the amplitude decay rate, $t$ is time.
The evolution of the initial states Eq.~(\ref{BState}) under this quantum channel can be described by
\begin{equation}
  \begin{split}
     \mathcal{E}_a (\rho) &= K_{1a} \rho K_{1a}^{\dagger} + K_{2a} \rho K_{2a}^{\dagger} \\
                         &= \frac{1}{4} \left(
                         \begin{array}{cccc}
                           \eta^2 (1+c_3) &       0        &          0          &    \eta(c_1-c_2)    \\
                                 0        & \eta^2 (1-c_3) &    \eta(c_1+c_2)    &          0          \\
                                 0        & \eta (c_1+c_2) & 2-\eta^2-\eta^2 c_3 &          0          \\
                           \eta (c_1-c_2) &       0        &          0          & 2-\eta^2+\eta^2 c_3
                         \end{array}
                         \right) \\
                         &= \frac{1}{4} (\mathbb{1}\otimes\mathbb{1} + r(t)\sigma_3\otimes\mathbb{1} + s(t)\mathbb{1}\otimes\sigma_3 +\sum_{j=1}^{3} c_j(t) \sigma_j\otimes\sigma_j), 
  \end{split}
\end{equation}
where $r(t) = \eta^2 -1$, $s(t) = 0$, $c_1(t) = \eta c_1$, $c_2(t) = \eta c_2$ and $c_3(t) = \eta^2 c_3$.
By Eq.~(\ref{IP3}), the IP of $\mathcal{E}_a (rho)$ is
\begin{equation}
  \mathcal{IP}(\mathcal{E}_a (\rho)) = \left\{ 
    \begin{array}{cc}
      \min\{ M_{22}(t),M_{33}(t) \}, & \text{if} |c_1(t)| < |c_2(t)|, \\
      \min\{ M_{11}(t),M_{33}(t) \}, & \text{if} |c_1(t)| > |c_2(t)|.
    \end{array}
\right.
\end{equation}
In particular, Eq.~(\ref{M123}) can reduce to the following simple form when $t = 0$:
\begin{equation}\label{MBell}
  \begin{split}
     M_{11}(t) &= \frac{c_2^2 + c_3^3 + 2c_1 c_2 c_3}{1-c_1^2}, \\
     M_{22}(t) &= \frac{c_1^2 + c_3^3 + 2c_1 c_2 c_3}{1-c_2^2}, \\
     M_{33}(t) &= \frac{c_2^2 + c_2^3 + 2c_1 c_2 c_3}{1-c_3^2}.
  \end{split}
\end{equation}

\begin{figure}[t]
  \centering
  \subfigure[$c_1 = 0.4, c_2 = 0.2, c_3 = 0.3$]{
    \includegraphics[width=0.45\textwidth]{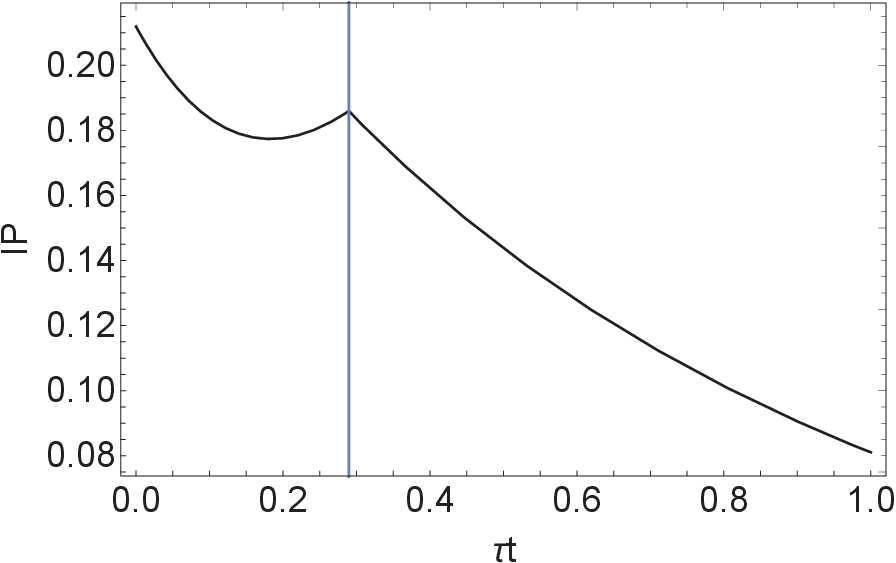}
  } 
  \subfigure[$c_1 = 0.4, c_2 = 0.2, c_3 = 0.3$]{
    \includegraphics[width=0.45\textwidth]{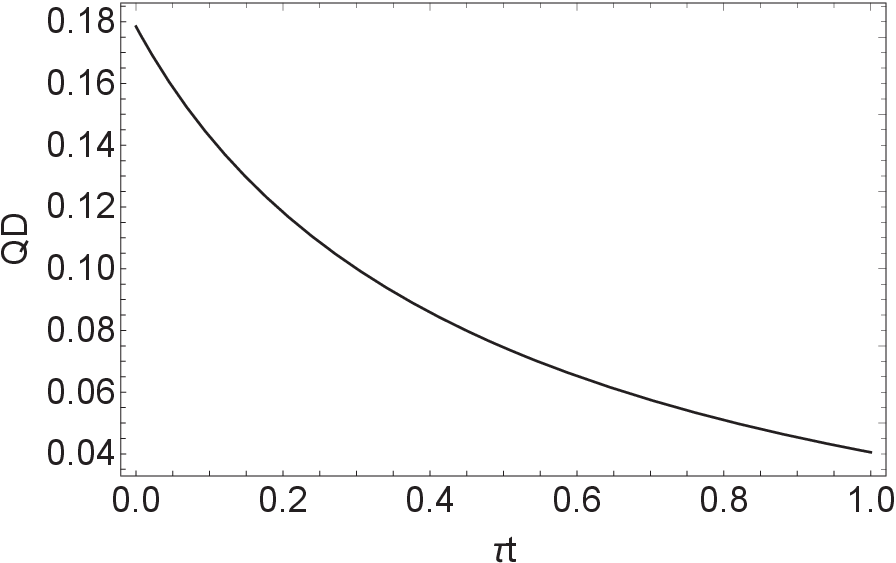}
  } \\
  \subfigure[$c_1 = 0.3, c_2 = 0.2, c_3 = 0.301$]{
    \includegraphics[width=0.45\textwidth]{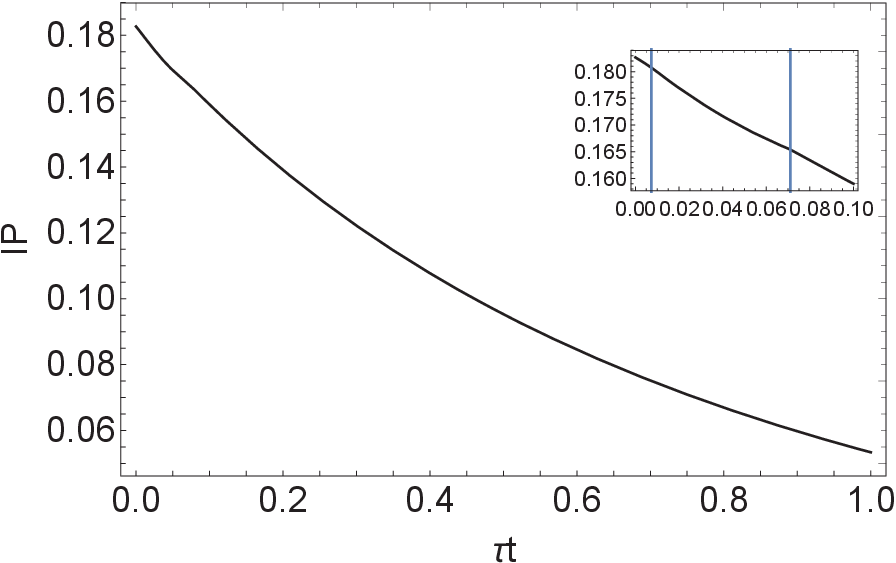}
  } 
  \subfigure[$c_1 = 0.3, c_2 = 0.2, c_3 = 0.301$]{
    \includegraphics[width=0.45\textwidth]{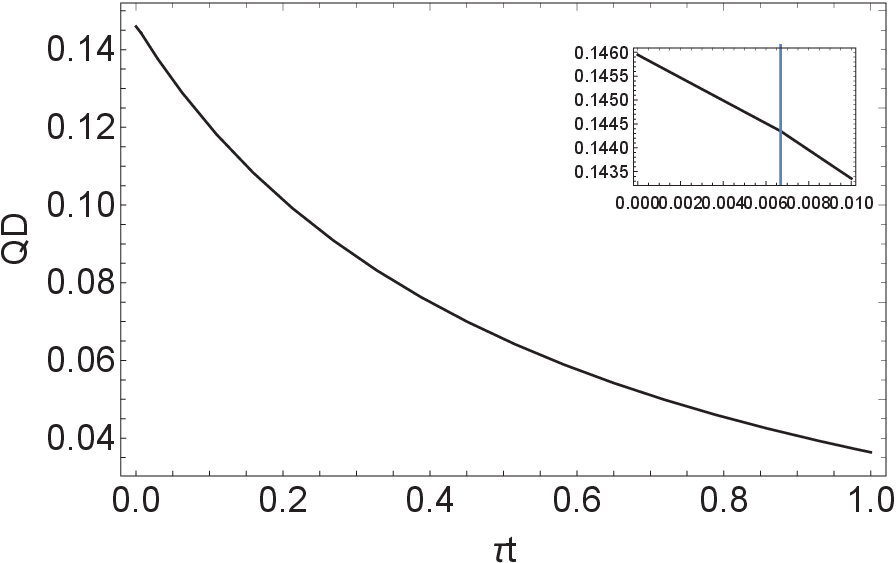}
  } 
  \caption{The evolution of IP ((a) and (c)) versus the evolution of quantum discord ((b) and (d)) under amplitude noise acting on the first qubit of the bipartite system.
  Considering the initial state not satisfying the condition of a sudden change of quantum discord ($c_1 = 0.4, c_2 = 0.2, c_3 = 0.3$), we can see that IP exhibits a sudden change with the same situation.
  On the other hand, if we consider the initial state satisfying the condition of a sudden change of quantum discord ($c_1 = 0.3, c_2 = 0.2, c_3 = 0.301$), one can see that a sudden change of IP still occurs with the same situation. 
   }\label{Fig1}
\end{figure}

It is easy to find that $M_{11}(0) > M_{22}(0)$ (or $M_{11}(t) < M_{22}(0)$) if $|c_1| < |c_2|$ (or $|c_1| > |c_2|$), and
when $t\rightarrow \infty$, $M_{11}(\infty) = M_{22}(\infty) > M_{33}(\infty)$.
Hence, a sudden change of IP occurs when $|c_3| < \max\{|c_1|,|c_2|\}$ in the initial Bell-diagonal state $\rho$ with this kind of amplitude noisy channel.
However, a sudden change of quantum discord occurs when $|c_3| > \max\{|c_1|,|c_2|\}$ in the initial state $\rho$ with the same noise environment.
We find that the initial state satisfying the condition of a sudden change of quantum discord also exhibits a sudden change of IP, which means that the conditions of the sudden change of IP and quantum discord are not complementary.
These results have been shown in Fig.~\ref{Fig1}.

\subsection{Phase noise}

Next we investigate the dynamics of IP over phase noise channel.
Phase noise describes the loss of quantum information without loss of energy.
The Kraus operators of phase noise for the whole system can be read as\cite{Nielsen2000,PRA78022322}
\begin{equation}
  \begin{split}
     K_{1p} &=\sqrt{\alpha} \left(
     \begin{array}{cc}
         1 & 0 \\
         0 & 1 
     \end{array}
      \right) \otimes
      \left(
     \begin{array}{cc}
         1 & 0 \\
         0 & 1 
     \end{array}
      \right), \\
     K_{2p} &=\sqrt{1-\alpha} \left(
     \begin{array}{cc}
       1 &  0 \\
       0 & -1 
     \end{array}
      \right) \otimes
      \left(
     \begin{array}{cc}
         1 & 0 \\
         0 & 1 
     \end{array}
      \right), \\
  \end{split}
\end{equation}
where $\alpha = \frac{1}{2}(1+ \sqrt{1-\gamma})$, $\gamma = e^{-\frac{\tau t}{2}}$ and $\tau$ denotes transversal decay rate.
The output states from this quantum channel is
\begin{equation}
  \begin{split}
     \mathcal{E}_p (\rho) &= K_{1p} \rho K_{1p}^{\dagger} + K_{2p} \rho K_{2p}^{\dagger} \\
                         &= \frac{1}{4} \left(
                         \begin{array}{cccc}
                                1 + c_3    &        0        &        0        & \gamma(c_1-c_2)  \\
                                  0        &      1 - c_3    & \gamma(c_1+c_2) &        0         \\
                                  0        & \gamma(c_1+c_2) &      1 - c_3    &        0         \\
                           \gamma(c_1-c_2) &        0        &        0        &      1 + c_3
                         \end{array}
                         \right) \\
                         &= \frac{1}{4} (\mathbb{1}\otimes\mathbb{1} +\sum_{i=1}^{3} c_i(t) \sigma_i\otimes\sigma_i), 
  \end{split}
\end{equation}
where $c_1(t) = \gamma c_1$, $c_2(t) = \gamma c_2$ and $c_3(t) = c_3$.
By Eq.~(\ref{IP3}), the expression of IP of $\mathcal{E}_p (\rho)$ can be obtained, that is
\begin{equation}
  \mathcal{IP}(\mathcal{E}_p (\rho)) = \left\{ 
    \begin{array}{cc}
      \min\{ M_{22}(t),M_{33}(t) \}, & \text{if} |c_1(t)| < |c_2(t)|, \\
      \min\{ M_{11}(t),M_{33}(t) \}, & \text{if} |c_1(t)| > |c_2(t)|.
    \end{array}
\right.
\end{equation}
\begin{figure}
  \centering
  \includegraphics[width=1\textwidth]{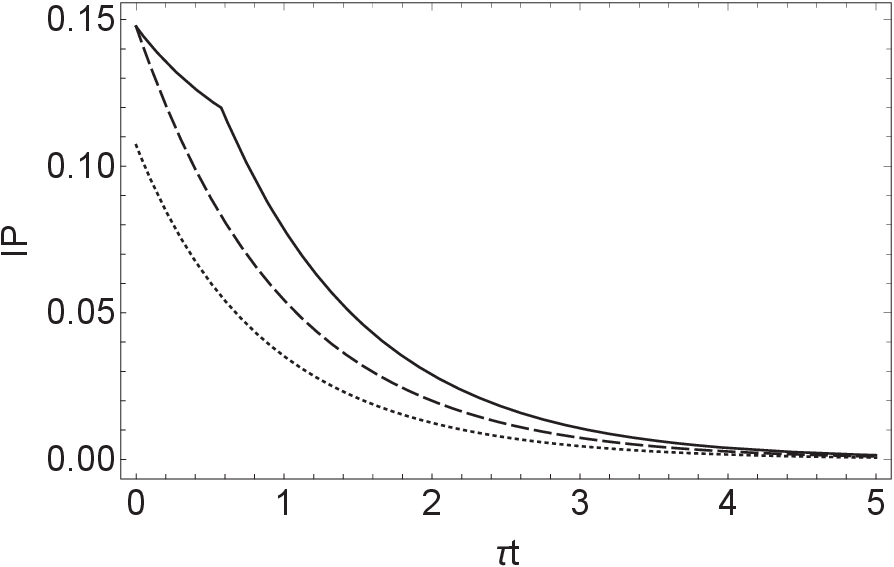}
  \caption{IP of the initial Bell-diagonal states under phase noise acting on the first qubit of the quantum system. 
  (1) $c_1 = 0.4, c_2 = 0.1, c_3 = 0.3$ (solid line).
  (2) $c_1 = 0.1, c_2 = 0.3, c_3 = 0.4$ (dashed line).
  (3) $c_1 = 0.4, c_2 = 0.3, c_3 = 0$ (dotted line).
  The sudden change only happens at situation (1).}\label{Fig2}
\end{figure}
where
\begin{equation}
  \begin{split}
     M_{11}(t) &=\dfrac{\gamma^2 c_2^2 +c_3^2 +2\gamma^2 c_1 c_2 c_3}{1-\gamma^2 c_1^2}, \\
     M_{22}(t) &=\dfrac{\gamma^2 c_1^2 +c_3^2 +2\gamma^2 c_1 c_2 c_3}{1-\gamma^2 c_2^2}, \\
     M_{33}(t) &=\dfrac{\gamma^2 c_1^2 +\gamma^2 c_2^2 +2\gamma^2 c_1 c_2 c_3}{1- c_3^2}.
  \end{split}
\end{equation}
One can find that the IP of $\mathcal{E}_p (\rho)$ can be rewritten as the following form:
\begin{equation} \label{BellIP}
  \mathcal{IP}(\mathcal{E}_p (\rho)) = \dfrac{\|C(t)\|^2 - \|C(t)\|_{\infty}^2 + 2\det C(t)}{1 - \|C(t)\|_{\infty}^2},
\end{equation}
where $\|C(t)\|^2 = \mathsf{Tr} [C^T C] = c_1^2(t) + c_2^2(t) + c_3^2(t)$ the square Hilbert-Schmidt of $C(t)$ and $\|C(t)\|_{\infty}^2 = \max\{ c_1^2(t), c_2^2(t), c_3^2(t) \}$ the operator norm of $C(t)$.\cite{PRL112210401}
If $|c_3| \geq \max\{ |c_1|, |c_2| \}$, $\|C(t)\|_{\infty}^2$ will reduce to $c_3^2(t)$ since the decay rate of $\max\{ |c_1(t)|, |c_2(t)| \}$ and $|c_3(t)|$ is different, then the IP $\mathcal{IP}(\mathcal{E}_p (\rho))$ decays monotonically.
If $|c_3| < \max\{ |c_1|, |c_2| \}$ and $c_3 \ne 0$, the decay rate of IP $\mathcal{IP}(\mathcal{E}_p (\rho))$ sudden changes at
$t_0 = -\frac{2}{\tau} \ln \left| \frac{c_3}{\max\{ |c_1|, |c_2| \}} \right|$.
In Fig.~\ref{Fig2}, we choose three different $\{c_i\}$ to represent the dynamics of IP of $\rho$ under single qubit phase noise.
As a result, a sudden change of IP may occur under this kind of noise environment.

\subsection{Depolarizing noise}

\begin{figure}[bp]
  \centering
  \includegraphics[width=1\textwidth]{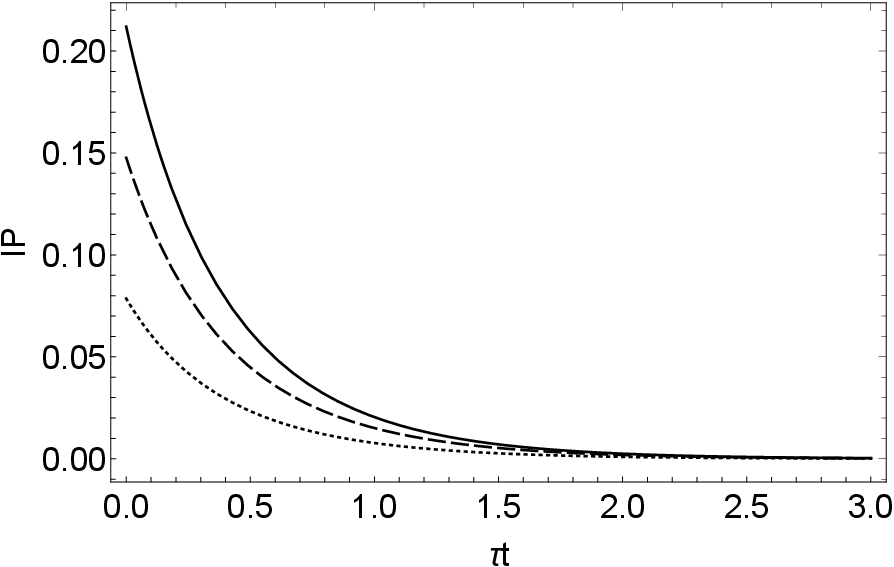}
  \caption{IP of the initial Bell-diagonal states under depolarizing noise acting on the first qubit of the quantum system. 
  (1) $c_1 = 0.4, c_2 = 0.3, c_3 = 0.2$ (solid line).
  (2) $c_1 = 0.1, c_2 = 0.3, c_3 = 0.4$ (dashed line).
  (3) $c_1 = 0.1, c_2 = 0.4, c_3 = 0.2$ (dotted line).
  }\label{Fig3}
\end{figure}

As one of the important types of quantum noise, the depolarizing noise describes a process that takes a state into completely mixed state $\mathbb{1}/2$ with probability $p$ and the state being left untouched with probability $1-p$.
The operation elements for depolarizing noise are shown as $\{ \sqrt{1-3p/4}\mathbb{1}, \sqrt{p}\sigma_x/2, \sqrt{p}\sigma_y/2, \sqrt{p}\sigma_z/2 \}$ and the Kraus operators for the whole system are given by\cite{Nielsen2000}
\begin{equation}
  \begin{split}
     K_{1d} &=\sqrt{1-\frac{3p}{4}} \left(
     \begin{array}{cc}
         1 & 0 \\
         0 & 1 
     \end{array}
      \right) \otimes
      \left(
     \begin{array}{cc}
         1 & 0 \\
         0 & 1 
     \end{array}
      \right), \ \ 
      K_{2d} =\frac{\sqrt{p}}{2} \left(
     \begin{array}{cc}
         0 & 1 \\
         1 & 0 
     \end{array}
      \right) \otimes
      \left(
     \begin{array}{cc}
         1 & 0 \\
         0 & 1 
     \end{array}
      \right),\\
      K_{3d} &=\frac{\sqrt{p}}{2} \left(
     \begin{array}{cc}
         0 & -i \\
         i & 0 
     \end{array}
      \right) \otimes
      \left(
     \begin{array}{cc}
         1 & 0 \\
         0 & 1 
     \end{array}
      \right), \ \ 
     K_{4d} =\frac{\sqrt{p}}{2} \left(
     \begin{array}{cc}
       1 &  0 \\
       0 & -1 
     \end{array}
      \right) \otimes
      \left(
     \begin{array}{cc}
         1 & 0 \\
         0 & 1 
     \end{array}
      \right), 
  \end{split}
\end{equation}
where $p=1-e^{-\tau t}$.
The Bell-diagonal states after this noise is
\begin{equation}
  \begin{split}
     \mathcal{E}_d (\rho) &=\!\sum_{j=1}^{4}\! K_{jd} \rho K_{jd}^{\dagger} \\
                         &=\! \frac{1}{4}\! \left(
                         \!\begin{array}{cccc} 
                            1 \!+\!(1\!-\!p) c_3  &           0            &            0           & (1\!-\!p)(c_1\!-\!c_2)  \\
                                     0            &  1 \!-\!(1\!-\!p) c_3  & (1\!-\!p)(c_1\!+\!c_2) &        0         \\
                                     0            & (1\!-\!p)(c_1\!+\!c_2) &  1 \!-\!(1\!-\!p) c_3  &        0         \\
                           (1\!-\!p)(c_1\!-\!c_2) &           0            &            0           &  1 \!+\!(1\!-\!p) c_3
                         \end{array}\!
                         \right) \\
                         &= \frac{1}{4} (\mathbb{1}\otimes\mathbb{1} +\sum_{i=1}^{3} c_i(t) \sigma_i\otimes\sigma_i), 
  \end{split}
\end{equation}
where $c_i(t) = (1-p)c_i$, $i=1,2,3$.
Similarly, the form of the IP of $\mathcal{E}_d (\rho)$ is Eq.~(\ref{BellIP}) but
$c_i(t) = (1-p)c_i$, $i=1,2,3$.
Since the identical decay rate of $|c_1(t)|$, $|c_2(t)|$ and $|c_3(t)|$, $\|C(t)\|_{\infty}^2$ will be determined once $\{ c_i \}$ has been chosen.
Therefore, there is no sudden change of IP of $\rho$ under depolarizing noise, see Fig.~\ref{Fig3}.

\section{Sudden change of IP under two-qubit dephasing model}

In this section, we investigate the dynamics of IP of Bell-diagonal states which are independently interacting with identical colored dephasing environment or are interacting with a common dephasing bath.

\subsection{Identical colored dephasing noise}

The study of the dynamics quantum open system based on the Markov approximation contains the majority of physical situation but lacking the case of system-environment interactions with memory.
In Ref.~\citen{PRA70010304}, a model describing system-environment interactions with memory without using Born-Markov approximation has been presented, and the conditions for system evolution to satisfy the complete positive trace-preserving map have been given.

The Kraus operators describing the above dynamics are given as follows:\cite{PRA88034304}
\begin{equation}
  M_1 = \sqrt{\beta} \left(
  \begin{array}{cc}
    1 & 0 \\
    0 & 1 
  \end{array}
  \right), \ \ 
  M_2 = \sqrt{1-\beta} \left(
  \begin{array}{cc}
    1 & 0 \\
    0 & -1 
  \end{array}
  \right),
\end{equation}
where the operators satisfy $\sum_{k}^{2} M_k^{\dagger} M_k =\mathbb{1}$, $\beta = \frac{1 + \Lambda(\nu)}{2}$ and $\Lambda(\nu) = e^{-\nu} [\cos(\mu\nu) + \sin(\mu\nu)/\mu]$,
$\mu = \sqrt{(4a\tau)^2-1}$, and $a$ is a coin-flip random variable, $\nu = \frac{t}{2\tau}$ is the dimensionless time.
After the evolution of Bell-diagonal states under the independent interaction with identical colored dephasing environment, the output density matrix can be obtained as
\begin{equation}
  \mathcal{E}_{ic} (\rho) = \sum_{i,j}(M_i \otimes M_j) \rho (M_i \otimes M_j)^{\dagger},
\end{equation}
where the operators $M_i$ and $M_j$ act on the first and second qubits, respectively.
Actually, this is a completely positive trace-preserving map.
\begin{figure}[htbp]
  \centering
  \includegraphics[width=1\textwidth]{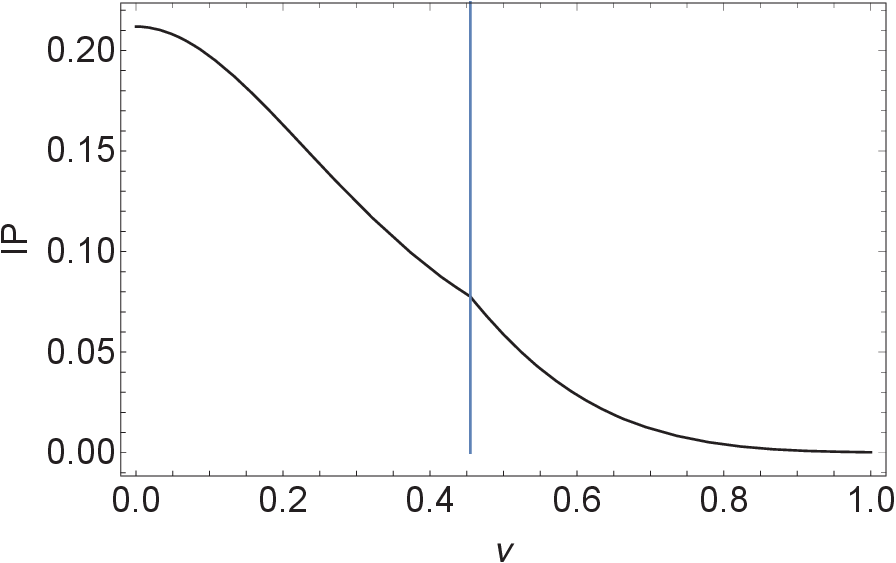}
  \caption{
  The dynamics of the IP of the initial Bell-diagonal state $\rho$, described by parameters $c_1=0.3, c_2=0.4, c_3=0.2$, independently interacting with identical colored dephasing noise having $a=1s$, $\tau=0.5s$.
  The decay rate of IP sudden changes at $\nu = 0.455$.
  }\label{Fig4}
\end{figure}
After the straightforward calculation, one can obtain that 
\begin{equation}
  \begin{split}
     \mathcal{E}_{ic} (\rho) &= \frac{1}{4} \left(
     \begin{array}{cccc}
                1+c_3          &            0            &            0            & \Lambda^2(\nu)(c_1-c_2) \\
                  0            &          1-c_3          & \Lambda^2(\nu)(c_1+c_2) &            0            \\
                  0            & \Lambda^2(\nu)(c_1+c_2) &          1-c_3          &            0            \\
       \Lambda^2(\nu)(c_1-c_2) &            0            &            0            &          1+c_3 
     \end{array}
     \right) \\
       &= \frac{1}{4} (\mathbb{1}\otimes\mathbb{1} + \sum_{i=1}^{3} c_i(t) \sigma_i \otimes \sigma_i),
  \end{split}
\end{equation}
where $c_1(t)= \Lambda^2(\nu) c_1$, $c_2(t) = \Lambda^2(\nu) c_2$ and $c_3(t) = c_3$.
By Eq.~(\ref{BellIP}), we get the form of the IP of $\mathcal{E}_{ic} (\rho)$.
Since $\Lambda^2(\nu)$ is monotonically decreasing as similar as $\gamma$, the sudden change of IP for $\mathcal{E}_{ic} (\rho)$ can occur with the proper $\{ c_i \}$, see Fig.~\ref{Fig4}.

\subsection{Two qubits coupling to common bath}

Considering the two-qubit system coupling to a same bosonic environment, the total Hamiltonian can be written as\cite{PRA97012126}
\begin{equation}
  H_c =\frac{1}{2} \sum_{j=1,2}\sigma_z^j + \sum_k \omega_k b_k^{\dagger} b_k + \sum_{j=1,2} \sum_k \sigma_z^j (g_k b_k^{\dagger} + \mathbf{H.c.}).
\end{equation}
\begin{figure}
  \centering
  \includegraphics[width=1\textwidth]{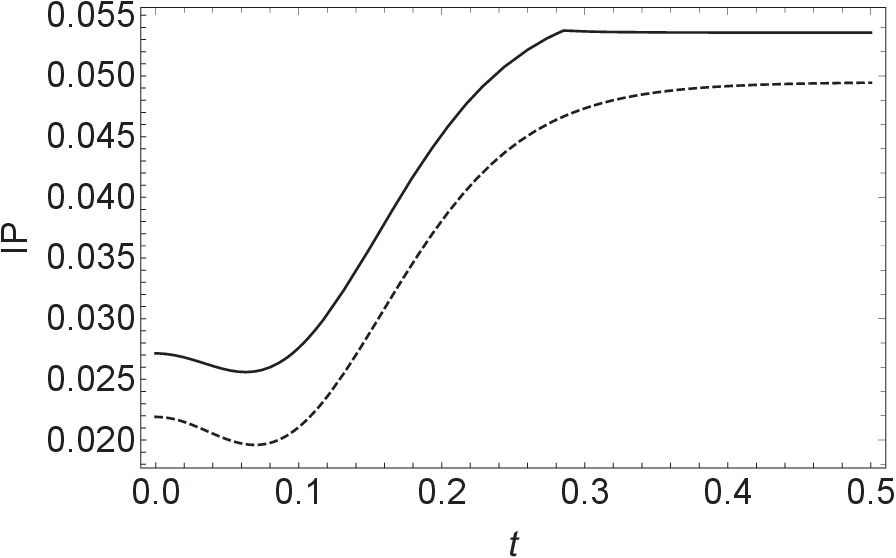}
  \caption{IP of the initial Bell-diagonal states under the two-qubit dephasing model with colored noise with 
  (1) $c_1=0.4, c_2=-0.1, c_3=0.16$ (solid line) and
  (2) $c_1=0.4, c_2=-0.1, c_3=0.14$ (dotted line), respectively.
  Sudden change of IP happens only at situation (2).
  }\label{Fig5}
\end{figure}
The dynamics of Bell-diagonal states under such common dephasing bath can be expressed by the following Kraus operators:
\begin{equation}
  K_{1cb} = \left(
  \begin{array}{cccc}
    \sqrt{\chi} &          0         &          0         &       0      \\
         0      & \frac{1}{\sqrt{2}} &          0         &       0      \\
         0      &          0         & \frac{1}{\sqrt{2}} &       0      \\
         0      &          0         &          0         & \sqrt{\chi} 
  \end{array}
  \right), \ \
  K_{2cb} = \left(
  \begin{array}{cccc}
    \sqrt{1-\chi} &          0         &          0         &        0       \\
          0       & \frac{1}{\sqrt{2}} &          0         &        0       \\
          0       &          0         & \frac{1}{\sqrt{2}} &        0       \\
          0       &          0         &          0         & -\sqrt{1-\chi} 
  \end{array}
  \right)
\end{equation}
where $\chi = \frac{\xi^4 + 1}{2}$, $\xi(t) = \exp[-\Gamma(t)]$, and $\Gamma(t)$ is the decoherence function with the form
\begin{equation}
  \Gamma(t) = \int_{0}^{\infty} \frac{1-\cos(\omega t)}{\omega^2} J(\omega) d(\omega).
\end{equation}
By considering the spectral density as $J(\omega) = \frac{\omega^s}{\omega_c^{s-1}} \exp(-\frac{\omega}{\omega_c})$ with the cut-off frequency $\omega_c =1 $ and the positive parameter $s=4$, we can investigate the dynamics of the Bell-diagonal states coupling to a same bosonic environment.

The matrix form of $\mathcal{E}_{cb}(\rho)$ with the initial Bell-diagonal state is as follow:
\begin{equation}
  \begin{split}
     \mathcal{E}_{cb}(\rho) &= K_{1cb}\rho K_{1cb}^{\dagger} + K_{2cb}\rho K_{2cb}^{\dagger} \\
       &= \frac{1}{4} \left(
       \begin{array}{cccc}
               1+c_3       &    0    &    0    & \xi^4(t)(c_1-c_2) \\
                 0         &  1-c_3  & c_1+c_2 &         0         \\
                 0         & c_1+c_2 &  1-c_3  &         0         \\
         \xi^4(t)(c_1-c_2) &    0    &    0    &       1+c_3 
       \end{array}
       \right) \\
       &= \frac{1}{4} (\mathbb{1}\otimes\mathbb{1} + \sum_{i=1}^{3} c_i(t) \sigma_i\otimes\sigma_i),
  \end{split}
\end{equation}
where $c_1(t) = [(1+\xi^4(t))c_1 + (1-\xi^4(t))c_2]/2$, $c_2(t) = [(1-\xi^4(t))c_1 + (1+\xi^4(t))c_2]/2$, and $c_3(t) = c_3$.
By Eq.~(\ref{BellIP}) with the proper parameter $s=4$, we find that if $c_1=c_2$, 
$c_1(t)$ and $c_2(t)$ will equal to $c_1$, then the IP of $\mathcal{E}_{cb}(\rho)$ will reduce to a constant.
If $c_1 \ne c_2$, the sudden change of IP happens when $c_3 > (c_1 + c_2)/2$ with this kind of noise environment. We show this in Fig.~\ref{Fig5}.

\section{Conclusion}

In summary, we have investigated the dynamics of IP of the X class of quantum states under several different kinds of noise channel, such as amplitude noise, phase noise and depolarizing noise acting only on one qubit of the quantum system, and two different types of phase noise acting on both two qubits of the quantum system.
Our results show that, as one of the discordlike measure, IP exhibits sudden change behavior, which is as similar as quantum discord.
Compared with dynamics of IP under depolarizing noise, the sudden change of IP occurs under amplitude noise and phase noise if chosen proper initial states.
In Ref.~\citen{IJQI111350048}, sudden change of quantum discord under one side quantum channel is shown.
In comparison, we show that sudden change of IP occurs when the quantum noise acts only on one qubit of the quantum system, which means that composite noise is not the necessary condition for the occurrence of sudden change of IP.
Furthermore, our results show that the initial state $\rho$ having no sudden change of quantum discord exhibits a sudden change of IP under the dynamics of amplitude noise, 
but the converse is not true.

\section{Acknowledgement}

F. L. Z. was supported by the National Natural Science Foundations of China (Grants Nos. 11675119). J. L. C. was supported by the National Natural Science Foundations of China (Grants Nos. 12275136 and 12075001). D. Z. was supported by the Nankai Zhide Foundations.

\end{document}